\documentclass[%
  reprint,
  superscriptaddress,
  showkeys,
  showpacs,
  nofootinbib,
  amsmath,amssymb,
  aps,
  prl,
  floatfix,
]{revtex4-1}
\usepackage{graphicx}
\usepackage{dcolumn}
\usepackage{bm}

\begin{document}
  \newcommand {\nc} {\newcommand}
  \nc {\IR} [1]{\textcolor{red}{#1}}

\preprint{APS/123-QED}


\title{ Halo nucleus $^{11}$Be:  a spectroscopic study via neutron transfer}

\author{K.T. Schmitt}%
\affiliation{Department of Physics and Astronomy, University of Tennessee, Knoxville, Tennessee 37996, USA}%
\affiliation{Physics Division, Oak Ridge National Laboratory, Oak Ridge, Tennessee 37831, USA}%
\author{K.L Jones}
\affiliation{Department of Physics and Astronomy, University of Tennessee, Knoxville, Tennessee 37996, USA}
\author{A. Bey}
\affiliation{Department of Physics and Astronomy, University of Tennessee, Knoxville, Tennessee 37996, USA}
\author{S.H. Ahn}
\affiliation{Department of Physics and Astronomy, University of Tennessee, Knoxville, Tennessee 37996, USA}
\author{D.W. Bardayan}%
\affiliation{Physics Division, Oak Ridge National Laboratory, Oak Ridge, Tennessee 37831, USA}%
\author{J.C. Blackmon}%
\affiliation{Department of Physics and Astronomy, Louisiana State University, Baton Rouge, Louisiana 70803-4001 USA}%
\author{S.M.~Brown}
\affiliation{Department of Physics, University of Surrey, Guildford, Surrey, GU2 7XH, UK}
\author{K.Y.~Chae}%
\altaffiliation{Physics Division, Oak Ridge National Laboratory, Oak Ridge, Tennessee 37831, USA}%
\altaffiliation{Department of Physics and Astronomy, University of Tennessee, Knoxville, Tennessee 37996, USA}
\affiliation{Department of Physics, Sungkyunkwan University, Suwon 440-746, Korea}
\author{K.A.~Chipps}%
\affiliation{Physics Department, Colorado School of Mines, Golden, Colorado 80401, USA}%
\author{J.A.~Cizewski}
\affiliation{Department of Physics and Astronomy, Rutgers University, New Brunswick, New Jersey 08903, USA}
\author{K. I.~Hahn}
\affiliation{Department of Science Education, Ewha Womans University, Seoul 120-750, Korea}
\author{J.J.~Kolata}
\affiliation{Department of Physics, University of Notre Dame, Notre Dame, Indiana 46556, USA}
\author{R.L.~Kozub}
\affiliation{Department of Physics, Tennessee Technological University, Cookeville, Tennessee 38505, USA}
\author{J.F.~Liang}
\affiliation{Physics Division, Oak Ridge National Laboratory, Oak Ridge, Tennessee 37831, USA}%
\author{C.~Matei}
\altaffiliation[Current address ]{EC-JRC - Institute for Reference Materials and Measurements, B-2440 Geel, Belgium}
\affiliation{Physics Division, Oak Ridge National Laboratory, Oak Ridge, Tennessee 37831, USA}
\author{M.~Mato\v{s}}
\affiliation{Department of Physics and Astronomy, Louisiana State University, Baton Rouge, Louisiana 70803-4001 USA}
\author{D.~Matyas}
\affiliation{Department of Physics and Astronomy, Denison University, Granville, Ohio 43023, USA}
\author{B.~Moazen}
\affiliation{Department of Physics and Astronomy, University of Tennessee, Knoxville, Tennessee 37996, USA}
\author{C.~Nesaraja}
\affiliation{Physics Division, Oak Ridge National Laboratory, Oak Ridge, Tennessee 37831, USA}%
\author{F.M.~Nunes}
\affiliation{National Superconducting Cyclotron Laboratory and Department of Physics and Astronomy, Michigan State University, East Lansing, Michigan 48824, USA}
\author{P.D.~O'Malley}
\affiliation{Department of Physics and Astronomy, Rutgers University, New Brunswick, New Jersey 08903, USA}
\author{S.D.~Pain}
\affiliation{Physics Division, Oak Ridge National Laboratory, Oak Ridge, Tennessee 37831, USA}%
\author{W.A.~Peters}
\affiliation{Department of Physics and Astronomy, Rutgers University, New Brunswick, New Jersey 08903, USA}
\author{S.T.~Pittman}
\affiliation{Department of Physics and Astronomy, University of Tennessee, Knoxville, Tennessee 37996, USA}
\author{A.~Roberts}
\affiliation{Department of Physics, University of Notre Dame, Notre Dame, Indiana 46556, USA}
\author{D.~Shapira}
\affiliation{Physics Division, Oak Ridge National Laboratory, Oak Ridge, Tennessee 37831, USA}%
\author{J.F.~Shriner~Jr.}
\affiliation{Department of Physics, Tennessee Technological University, Cookeville, Tennessee 38505, USA}
\author{M.S.~Smith}
\affiliation{Physics Division, Oak Ridge National Laboratory, Oak Ridge, Tennessee 37831, USA}%
\author{I.~Spassova}
\affiliation{Department of Physics and Astronomy, Rutgers University, New Brunswick, New Jersey 08903, USA}
\author{D.W.~Stracener}
\affiliation{Physics Division, Oak Ridge National Laboratory, Oak Ridge, Tennessee 37831, USA}%
\author{A.N.~Villano}
\affiliation{Physics Department, University of Michigan, Ann Arbor, Michigan 48109, USA}
\author{G.L.~Wilson}
\affiliation{Department of Physics, University of Surrey, Guildford, Surrey, GU2 7XH, UK}

\date{\today}

\begin{abstract}

The best examples of halo nuclei, exotic systems with a diffuse nuclear cloud surrounding a tightly-bound core, are found in the light, neutron-rich region, where the halo neutrons experience only weak binding and a weak, or no, potential barrier.  Modern direct reaction measurement techniques provide powerful probes of the structure of exotic nuclei.  Despite more than four decades of these studies on the benchmark one-neutron halo nucleus $^{11}$Be, the spectroscopic factors for the two bound states remain poorly constrained.  In the present work, the $^{10}$Be($d$,$p$) reaction has been used in inverse kinematics at four beam energies to study the structure of $^{11}$Be.    The spectroscopic factors extracted using the adiabatic model, were found to be consistent across the four measurements, and were largely insensitive to the optical potential used.   The extracted spectroscopic factor for a neutron in a $n\ell j=2s_{1/2}$ state coupled to the ground state of $^{10}$Be is 0.71(5).  For the first excited state at 0.32 MeV, a spectroscopic factor of 0.62(4) is found for the halo neutron in a 1$p_{1/2}$ state.

\end{abstract}

\maketitle


Nuclear halos are a phenomenon associated with certain weakly-bound nuclei, in which a tail of dilute nuclear matter is distributed around a tightly bound core \cite{tan85,Alk96,Jen04}. This effect is only possible for bound states with no strong Coulomb or centrifugal barrier, and which lie close to a particle-emission threshold. Though excited-state halos exist, the number of well-studied halo states is predominantly limited to a handful of light, weakly-bound nuclei which exhibit the phenomenon in their ground state.

The neutron-rich nucleus $^{11}$Be is a brilliant example of this phenomenon, with halo structures in both of its bound states, and light enough to be modeled with an \emph{ab initio} approach.  It is well documented that the 1/2$^+$ ground state and 1/2$^-$ first excited state in $^{11}$Be are inverted with respect to level ordering predicted from a na\"{i}ve shell model.  There has been considerable theoretical effort toward reproducing this level inversion in a systematic manner, while maintaining the standard ordering in the nearby nuclide $^{13}$C, where the 1/2$^+$ state lies over 3 MeV above the 1/2$^-$ ground state. A Variational Shell Model approach \cite{Ots93} and models which vary the single-particle energies via vibrational \cite{Vin95} and rotational \cite{Nun96} core couplings reproduce this level inversion in a systematic manner. Common to the success of these models is the inclusion of core excitation. \emph{Ab initio} No-Core Shell Model calculations \cite{For05} have been unable to reproduce this level inversion though a significant drop in the energy of the 1/2$^+$ state in $^{11}$Be is reported with increasing model space. In all of these models, the wave functions for the $^{11}$Be halo states show a considerable overlap with a valence neutron coupled to an excited $^{10}$Be($2^+$) core, in addition to the na\"{i}ve $n\otimes^{10}$Be($0_{\mathrm{gs}}^+$) component.  Despite decades of study, the extent of this mixing is not well understood, with both structure calculations and the interpretation of experimental results ranging from a few percent to over 50 percent core-excited component. 

\begin{figure}
\includegraphics[width=6.5cm]{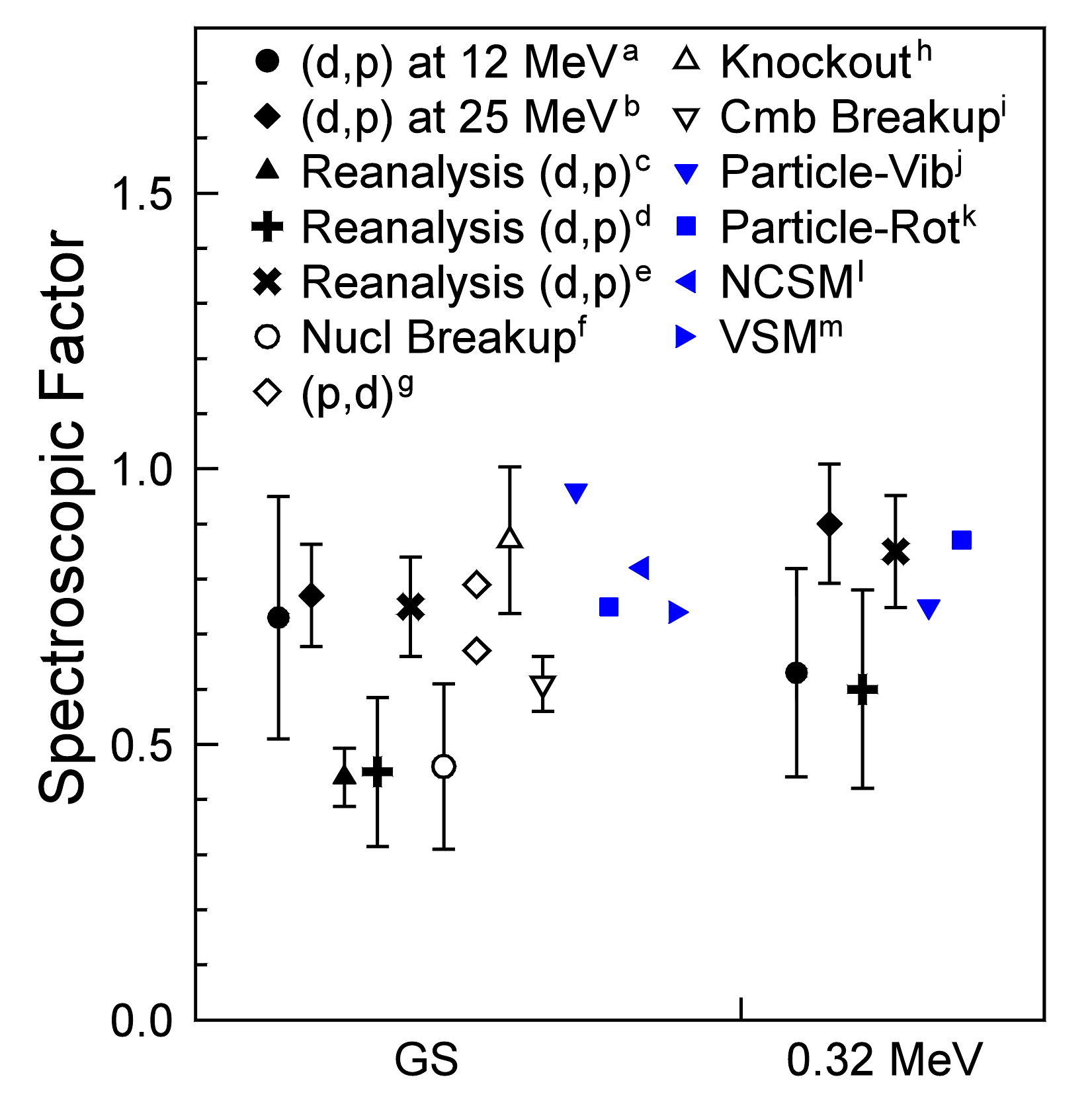}
\caption{\label{fig1}A selection of literature values for spectroscopic factors for the ground and first excited states of $^{11}$Be.  Experimentally extracted spectroscopic factors are taken from references  \cite{aut70}(a), \cite{zwi79}(b), \cite{Tim99}(c), \cite{del09}(d,e), \cite{Lim07}(f), \cite{Win01}(g), \cite{Aum00}(h), and \cite{Pal03}(i).  Results of theoretical calculations are taken from  \cite{Vin95}(j), \cite{Nun96}(k), \cite{For05}(l), and \cite{Ots93}(m).  Filled black symbols denote spectroscopic factors extracted from ($d$,$p$) data.\protect\\}
\end{figure}
Experimentally, it is possible to gain quantitative insight into the mixed configuration of a state by studying reactions which provide observables that are sensitive to different components of the nuclear wave function. By comparing the measured differential cross sections to those calculated theoretically, a spectroscopic factor $S$ can be extracted, which reflects in a model-dependent way the extent to which the nuclear state studied resembles that used in the calculation. Spectroscopic factors reported from numerous direct-reaction studies, including one-neutron transfer \cite{aut70,zwi79,for99,Win01}, two-neutron transfer \cite{liu90}, neutron knockout \cite{Aum00}, Coulomb breakup \cite{Pal03}, and reanalyses of the neutron-transfer data \cite{Tim99, del09}, are shown in Figure \ref{fig1}, along with those from theoretical calculations. For the ground state, the spectroscopic factors correspond to overlaps with a $2s_{1/2}\otimes ^{10}$Be(0$_{\mathrm{gs}}^+$) configuration. For the first excited state, they refer to the $1p_{1/2}\otimes^{10}$Be(0$_{\mathrm{gs}}^+$). These values exhibit significant variation beyond the reported uncertainties, even within the same reaction type.

The challenges associated with determining spectroscopic factors from measurements using various reactions has received considerable attention in recent years \cite{Kra01,Gad08}.  Even extracting consistent spectroscopic factors from a single reaction type is sensitive to the description of the reaction mechanism (evidenced by the variation in $S$ for different analyses of ($d$,$p$) data), and to the nuclear structure included in the model (evidenced by the range of $S$ from ($p$,$d$) studies \cite{for99,Win01} using different potentials to calculate the wave functions). However, it should be expected that consistent results are obtained for measurements employing the same reaction type, provided the data are analyzed in the same manner with a realistic reaction description.

Measured cross sections for transfer reactions, such as ($d$,$p$), have traditionally been analyzed by assuming that the reaction occurs in a single step, so that the degree of configuration mixing can be gauged by comparing measured differential cross sections to those calculated using the Distorted Wave Born Approximation (DWBA). A spectroscopic factor is extracted, which for ($d$,$p$) indicates the extent to which the final state resembles the ground state of the core coupled to a neutron in a particular single-particle state. This formalism relies on nuclear scattering potentials to describe the interaction between the two nuclei in the entrance and exit channels. These potentials are traditionally constrained by measurements of elastic scattering. In the case of reactions on unstable nuclei, the elastic scattering data do not always exist, leading to ambiguities in the DWBA analysis. Additionally, as the deuteron is relatively loosely bound (by only 2.2~MeV), it may break up in the presence of the target nucleus.  This deuteron breakup channel can couple to the transfer channel, affecting the spectroscopic factors extracted in a non-trivial manner. 

To account for this mechanism, Johnson and Soper \cite{Joh70} devised the ADiabatic Wave Approximation (ADWA), which uses nucleonic potentials and explicitly includes deuteron breakup.  An extension of this method to include finite range effects (ADWA-FR) was introduced by Johnson and Tandy \cite{Joh74}.  The validity of FR-ADWA has been demonstrated in \cite{Nun11}, where FR-ADWA was compared in detail to exact Faddeev calculations. Therein it is shown that including the first term of the Weinberg expansion works extremely well at these low energies and not as well at energies used in fragmentation facilities. 

The two sets of existing $^{10}$Be($d$,$p$)$^{11}$Be data, measured at $E_d = 12$ MeV \cite{aut70} and $E_d = 25$ MeV \cite{zwi79}, have been reanalyzed multiple times \cite{Tim99,kee04, del09} yielding widely differing $S$ values.  Even using consistent analyses for both sets of data did not result in compatible spectroscopic factors \cite{kee04,del09}. It is unclear to what extent these discrepancies in $S$ are due to experimental issues (such as the normalization of the two sets of data), or insufficiently constrained theoretical analysis.

To help resolve this situation, an extensive series of measurements have been made using a primary beam of $^{10}$Be ions. Elastic and inelastic scattering data were obtained simultaneously with neutron transfer at four energies, covering almost the entire energy range between the ($d$,$p$) measurements described in the literature. The data presented here are for the one-neutron transfer ($d$,$p$) reaction to bound states and deuteron elastic scattering; transfer to resonant states and further scattering data will be presented in future publications. The transfer data were analyzed using both ADWA-FR \cite{Joh74,Ngu10}, and with the traditional DWBA method.

Data were collected during two separate experimental campaigns at the Holifield Radioactive Ion Beam Facility \cite{Bee11} at Oak Ridge National Laboratory, using similar light-ion detection setups.  The initial run was performed at a $^{10}$Be beam energy of 107~MeV ($E_d = 21.4$~MeV), with the subsequent runs  at 60, 75, and 90 MeV ($E_d = 12$, 15, and 18~MeV), with approximately $5 \times 10^6$ particles per second.  The $^{10}$Be ions were accelerated from a cesium sputter ion source using the 25 MV tandem accelerator.   Contamination from $^{10}$B was reduced to less than 1\% by fully stripping the beam ions and tuning the energy-analyzing magnet for $Z$ = 4.  Deuterated polyethylene targets with areal densities of 94, 162, and 185~$\mu$g/cm$^2$ were used.  

 \begin{figure}
\centering
\includegraphics[width=8cm]{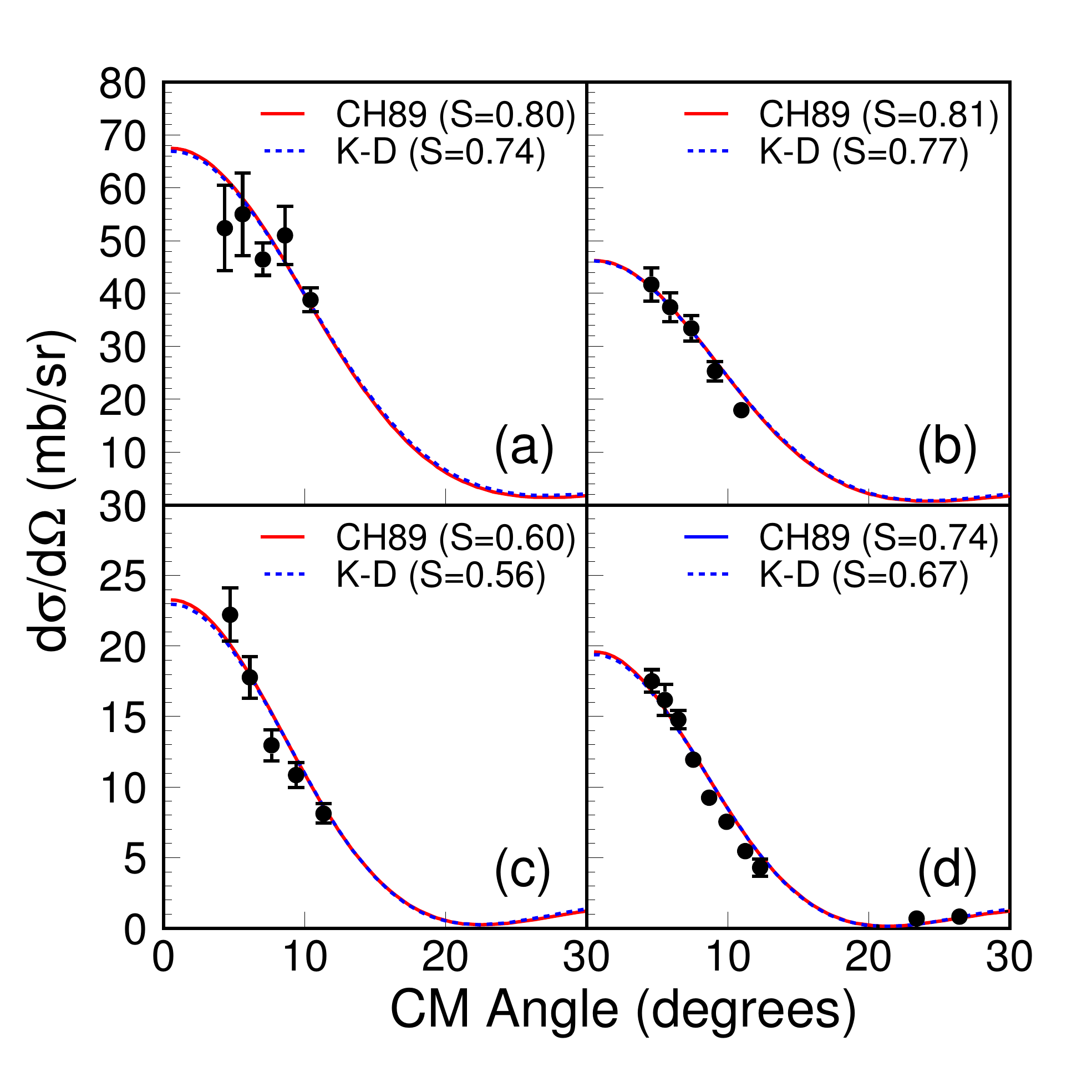}
\caption{Differential cross sections for transfer to the ground state of $^{11}$Be for equivalent deuteron energies of 12.0 (a), 15.0 (b), 18.0 (c), and 21.4 (d) MeV.  Cross sections were calculated using the ADWA-FR of Johnson and Tandy \cite{Joh74, Ngu10}, using the CH89 \cite{Var91} and K-D \cite{Kon03} optical potentials.  Calculated cross sections are scaled using the indicated spectroscopic factors.\label{transfer1}}
\end{figure}

The angles and energies of light-ion ejectiles were measured using the Silicon Detector Array (SIDAR \cite{Bar01})(covering $138^{\circ}<\theta_{\textrm{lab}}<165^{\circ}$), and the first full implementation of the Oak Ridge Rutgers University Barrel Array (ORRUBA \cite{Pai07})($45^{\circ}<\theta_{\textrm{lab}}<135^{\circ}$).  The ORRUBA position-sensitive silicon strip detectors (1000~$\mu$m thick) were mounted at a radius of 76~mm at laboratory angles forward of $95^{\circ}$ and 87~mm at more backward angles.  

Light ions emitted at forward laboratory angles were identified on the basis of their differential stopping power.  An angular resolution of better than 2$^{\circ}$ in polar angle was achieved.   For the purpose of normalization, the product of target areal density and integrated beam exposure was determined for the transfer data using the elastically scattered deuterons measured in the forward angle ORRUBA detectors, with reference to a low-intensity run where the beam particles were counted directly.  Protons from the ($d$,$p$) reaction were detected in the SIDAR array with an energy resolution of $\approx$~70~keV at all beam energies.  The energies of protons emitted from the ($d$,$p$) reaction at the lowest three beam energies were too small to be measured in ORRUBA.  However, proton angular distributions for both bound states were measured in ORRUBA at E$_d$~=~21.4~MeV with an energy resolution of $\approx$ 200~keV.  In the first run, beam particles were counted using the new Dual Micro-Channel Plate (DMCP) detector for heavy recoil detection.  A new fast ionization chamber, similar to that described in reference \cite{kim05}, was used in the later runs for beam particle counting and identification.  

Ground-state angular distributions of protons emitted from the $^{10}$Be($d$,$p$)$^{11}$Be reaction are compared to ADWA-FR predictions normalized to the data in Fig. \ref{transfer1}. Optical potentials from Varner (CH89) \cite{Var91} and Koning and Delaroche (K-D) \cite{Kon03} were used for both the entrance and exit channels.  No significant differences are found in the shapes of the calculated angular distributions.  Good agreement with experimental data is seen for both of the bound states, for all four energies. Additionally, the data were compared with DWBA calculations (not shown here), which described the shape of the angular distributions well.
All transfer calculations in this work were performed with FRESCO \cite{tho88}, and adiabatic potentials
were obtained with a modified version of TWOFNR \cite{twofnr}. A fixed standard radius and diffuseness $r = 1.25$~fm and $a = 0.65$~fm were used for the bound state.  The Reid interaction \cite{Rei68} was used to obtain the deuteron wave function and in the transfer operator.

\begin{figure*}
\includegraphics[width=15cm]{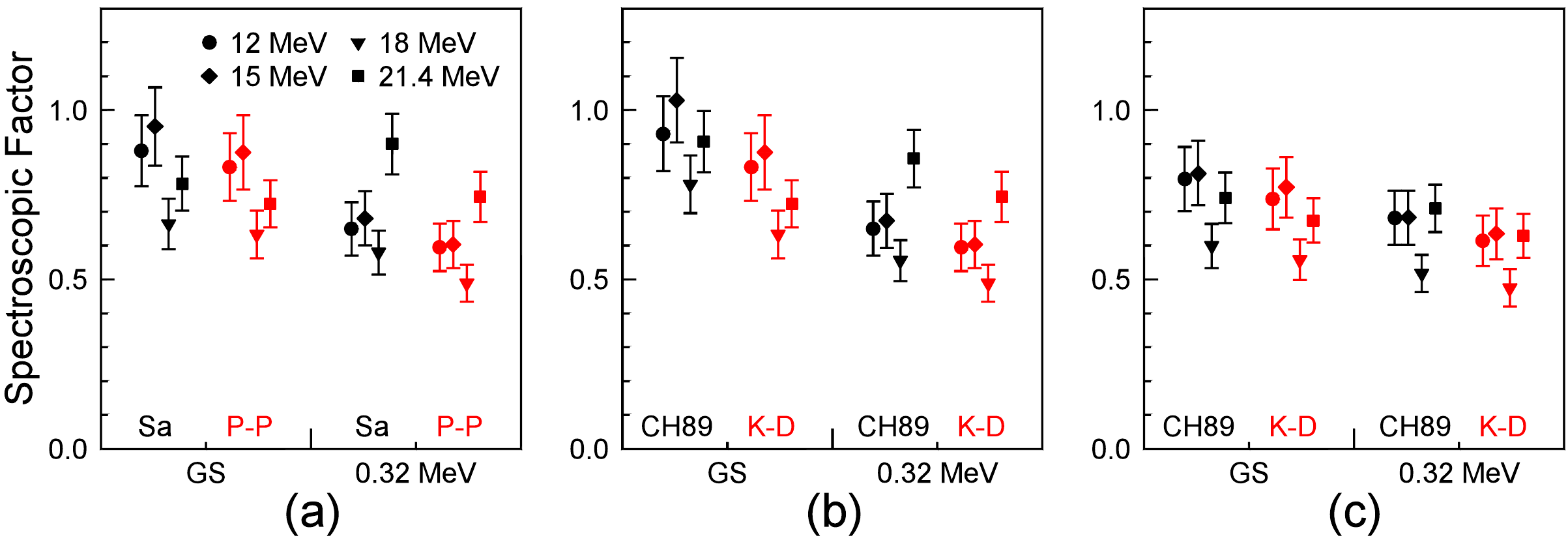}
\caption{\label{DWBA}(Color online) Spectroscopic factors for $^11$Be, extracted from ($d$,$p$) data with DWBA (panels (a) and (b)) and ADWA-FR (panel (c)) formalisms.  Panels a and b illustrate the effect of choosing different potentials in the entrance and exit channels, respectively.  The error bars are from experimental uncertainties only.   See text for further details.}
\end{figure*}

Spectroscopic factors were extracted for each state at each beam energy using both the DWBA and ADWA-FR formalisms.  
The results are shown in Fig.\ref{DWBA}:  Panel a shows the sensitivity to the deuteron optical potential (Satchler (Sa) \cite{Sat66}  versus Perey and Perey (P-P) \cite{Per63}, keeping the proton potential fixed (K-D) \cite{Kon03}); panel (b) shows the sensitivity to the proton potential in the exit channel (CH89 \cite{Var91} versus K-D with the deuteron potential P-P).  Spectroscopic factors for the ground (excited) state are shown on the left (right). 

\begin{figure}
\centering
\includegraphics[width=8.cm]{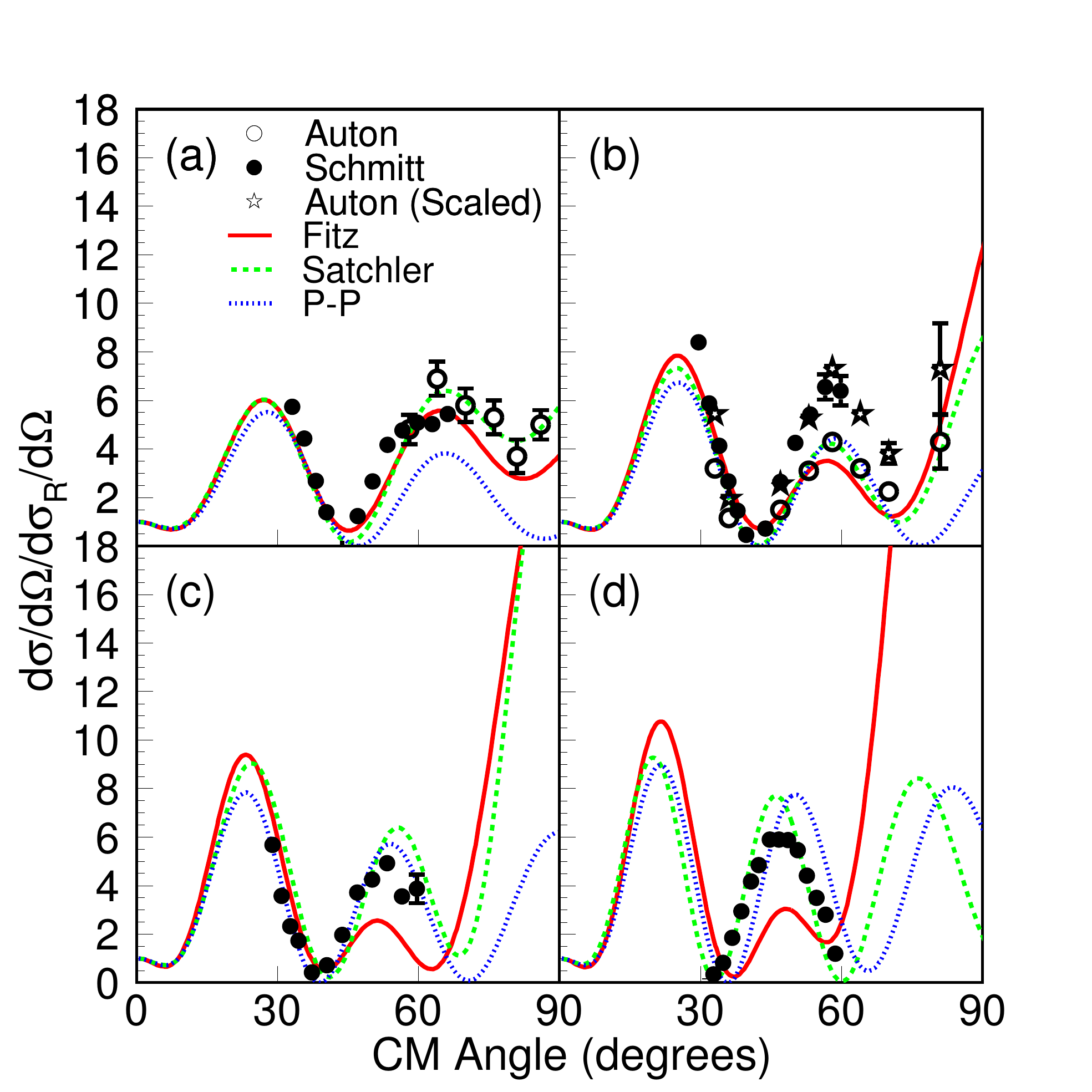}
\caption{(Color online) Elastic scattering differential cross-sections for equivalent deuteron energies of 12.0 (a), 15.0 (b), 18.0 (c), and 21.4 (d) MeV, including the data of Auton \cite{aut70} and the present study, and optical model calculations using the parameters of Fitz \cite{Fit67}, Satchler \cite{Sat66}, and Perey and Perey \cite{Per63}.\label{elastic}}
\end{figure}

The DWBA analysis is sensitive to the choice of optical potential and there is variation in the value of $S$ extracted at each of the four energies using the same optical potential.  This is most apparent at the highest beam energy for the first excited state.  These problems indicate shortcomings in the DWBA prescription (as discussed below).  In the case of the ADWA-FR analysis, only nucleon potentials are necessary; panel (c) of Fig. \ref{DWBA} shows the results obtained with CH89 versus K-D. In this case, the sensitivity to the chosen optical potential is reduced, and the $S$ extracted for the first excited state at the highest beam energy is brought into agreement with the results at lower energies.  The average $S$ extracted from our data are 0.71(5) for the ground state and 0.62(4) for the first excited state.
%

The inconsistencies arising in the DWBA analysis come in part from the deuteron optical potentials, as seen from elastic scattering. Fig.~\ref{elastic} shows the current elastic scattering data compared to those from Auton \cite{aut70}.  It should be noted that Auton normalized the data to optical model calculations of the deuteron elastic scattering using the Sa potential \cite{Sat66}.  In the current work, the elastic scattering data were analyzed using three deuteron optical potentials: Fitz \cite{Fit67}, Sa \cite{Sat66}, and P-P \cite{Per63}.  The potentials produce quite different angular distributions, and no calculation is able to adequately describe the elastic scattering data at all four energies.  This indicates that another mechanism, possibly breakup, is playing an important role which the deuteron optical potentials fail to describe in the $d+^{10}$Be reaction.  Furthermore, as the Auton data were normalized to one of these calculations, the spectroscopic factors extracted from those data are subject to a significant systematic uncertainty.  When the Auton deuteron elastic scattering data taken at E$_{beam}$~=~15~MeV are scaled to the present data (shown as stars in panel b of Fig.\ref{elastic}), excellent agreement is reached.   An alternate analysis of these data using a Continuum Discretized Coupled Channel approach has been completed and will be presented in a future publication. 

In summary, elastic scattering and neutron-transfer measurements, performed with a primary $^{10}$Be beam incident on deuterated plastic targets at E$_d$=~12 to 21.4~MeV, have been used to study the structure of $^{11}$Be halo states in a systematic manner. 
The poor reliability of deuteron optical potentials for these reactions is reflected in the scatter of the spectroscopic factor $S$ extracted from the transfer data and its dependence on the choice of optical potential.  Including deuteron breakup through ADWA-FR reduces both of these problems and results in a more reliable extraction of $S$. The average $S$ extracted from our data using the ADWA-FR formalism are 0.71(5) for the ground state and 0.62(4) for the first excited state.  The quoted uncertainties are solely from experimental considerations.  The improved energy-independence of the extracted S using the present data and analysis is in stark contrast to extractions from previous transfer data.
%
%








\begin{acknowledgments}
This work was supported by the US Department of Energy
under contract numbers DE-FG02-96ER40955 (TTU), DE-AC05-00OR22725 (ORNL),
DE-FG02-96ER40990 (TTU), DE-FG03-93ER40789 (CO School of Mines),
DE-FG02-96ER40983 and DE-SC0001174 (UT), DE-AC02-06CH11357
(MSU), and DE-SC0004087 (MSU), by the National Science Foundation under contract numbers
PHY0354870, PHY0757678 (Rutgers), PHY-1068571
(MSU), and PHY0969456 (Notre Dame), by the NRF of the MEST (Korea) under contract number 2010-0027136, and by the UK Science and Technology Facilities Council under contract
number PP/F000715/1.  This research was sponsored in part by the National Nuclear Security Administration
under the Stewardship Science Academic Alliance program through DOE Cooperative Agreement 
DE-FG52-08NA28552(Rutgers, ORAU, MSU).
\end{acknowledgments}

\bibliography{11Be_PRL}

\end{document}